\documentclass[prl,floatfix,twocolumn,superscriptaddress]{revtex4-2}
\usepackage{amsmath}
\usepackage{amssymb}
\usepackage{amstext}
\usepackage{amsopn}
\usepackage{amsfonts}
\usepackage{amsxtra}
\usepackage[english]{babel}
\usepackage{graphicx}
\usepackage{float}
\usepackage{bm}
\usepackage{multirow}
\usepackage{dcolumn}
\usepackage{color}
\usepackage{hyperref}

\newcommand{\icm}{\ensuremath{~\textrm{cm}^{-1}}}

\begin{document}

\title{\textbf{Optical Excitations of Flat Bands Induced by Exciton Condensation in Ta$_2$Pd$_3$Te$_{5}$}}

\author{Shaohui Yi}
\thanks{These authors contributed equally to this work.}
\affiliation{Beijing National Laboratory for Condensed Matter Physics, Institute of Physics, Chinese Academy of Sciences, P.O. Box 603, Beijing 100190, China}
\affiliation{School of Physical Sciences, University of Chinese Academy of Sciences, Beijing 100049, China}

\author{Zhiyu Liao}
\thanks{These authors contributed equally to this work.}
\affiliation{Beijing National Laboratory for Condensed Matter Physics, Institute of Physics, Chinese Academy of Sciences, P.O. Box 603, Beijing 100190, China}

\author{Chenhao Liang}
\thanks{These authors contributed equally to this work.}
\affiliation{Beijing National Laboratory for Condensed Matter Physics, Institute of Physics, Chinese Academy of Sciences, P.O. Box 603, Beijing 100190, China}
\affiliation{School of Physical Sciences, University of Chinese Academy of Sciences, Beijing 100049, China}

\author{Sheng Zhang}
\affiliation{Beijing National Laboratory for Condensed Matter Physics, Institute of Physics, Chinese Academy of Sciences, P.O. Box 603, Beijing 100190, China}
\affiliation{School of Physical Sciences, University of Chinese Academy of Sciences, Beijing 100049, China}

\author{Xiutong Deng}
\affiliation{Beijing National Laboratory for Condensed Matter Physics, Institute of Physics, Chinese Academy of Sciences, P.O. Box 603, Beijing 100190, China}
\affiliation{School of Physical Sciences, University of Chinese Academy of Sciences, Beijing 100049, China}
\author{Yongjie Xie}
\author{Lincong Zheng}
\author{Yujie Wang}
\author{Yubiao Wu}
\affiliation{Beijing National Laboratory for Condensed Matter Physics, Institute of Physics, Chinese Academy of Sciences, P.O. Box 603, Beijing 100190, China}

\author{Zhijun Wang}
\email[]{wzj@iphy.ac.cn}
\affiliation{Beijing National Laboratory for Condensed Matter Physics, Institute of Physics, Chinese Academy of Sciences, P.O. Box 603, Beijing 100190, China}
\affiliation{School of Physical Sciences, University of Chinese Academy of Sciences, Beijing 100049, China}

\author{Youguo Shi}
\email[]{ygshi@iphy.ac.cn}
\affiliation{Beijing National Laboratory for Condensed Matter Physics, Institute of Physics, Chinese Academy of Sciences, P.O. Box 603, Beijing 100190, China}
\affiliation{School of Physical Sciences, University of Chinese Academy of Sciences, Beijing 100049, China}

\author{Xianggang Qiu}
\email[]{xgqiu@iphy.ac.cn}
\affiliation{Beijing National Laboratory for Condensed Matter Physics, Institute of Physics, Chinese Academy of Sciences, P.O. Box 603, Beijing 100190, China}
\affiliation{School of Physical Sciences, University of Chinese Academy of Sciences, Beijing 100049, China}

\author{Bing Xu}
\email[]{bingxu@iphy.ac.cn}
\affiliation{Beijing National Laboratory for Condensed Matter Physics, Institute of Physics, Chinese Academy of Sciences, P.O. Box 603, Beijing 100190, China}
\affiliation{School of Physical Sciences, University of Chinese Academy of Sciences, Beijing 100049, China}

\date{\today}
%
%

\begin{abstract}
We report on the charge dynamics of Ta$_2$Pd$_3$Te$_5$ using temperature-dependent optical spectroscopy with polarized light. We observe a metal-insulator transition characterized by the collapse of Drude response and the emergence of sharp and narrow absorption peaks at low temperatures. Unlike previous excitonic insulator candidates such as TiSe$_2$ and Ta$_2$NiSe$_5$, where the excitonic order is intertwined with charge density wave or structural instabilities, the sharp features in Ta$_2$Pd$_3$Te$_5$ point to intrinsic excitonic excitations associated with ultra-flat bands driven by many-body renormalization of the band structure via spontaneous exciton condensation. Our findings thus provide clear-cut optical evidence for exciton condensation in a bulk crystal and establish Ta$_2$Pd$_3$Te$_5$ as a promising platform for exploring correlated quantum phases and novel excitonic phenomena.
\end{abstract}


\maketitle

%
%
Excitonic insulator, a quantum state arising from the spontaneous condensation of electron-hole pairs (excitons), was first proposed in the 1960s as a solid-state analog of superconductivity~\cite{Mott1961,Jerome1967,Kohn1967,Zittartz1967,HALPERIN1968}. This exotic state occurs when the exciton binding energy exceeds the bandgap of a semiconductor or semimetal, enabling the formation of a macroscopic coherent condensation. In this state, the condensation leads to a many-body renormalization of the band structure, flattening the valence band top and the conduction band bottom near the Fermi level and opening a collective insulating gap. Despite its intriguing concept, excitonic insulators are rarely found in real materials. Their realization requires several key conditions, including well-preserved particle-hole symmetry in electronic bands, low carrier density to minimize Coulomb screening, and sufficiently low temperatures to maintain the long-range coherence of excitonic states. Experimentally, most candidates have been identified in low-dimensional materials~\cite{Butov1994PRL,Naveh1996PRL,Eisenstein2004,Min2008PRL,Eisenstein2014,Du2017,Li2017,Wang2023,Kogar2017,Jauregui2019,Wang2019,Gupta2020NC,Ma2021,Sun2022NP,Jia2022NC,Gao2023,Song2023,Gao2024NC,Gao2024NP,Lian2024}, such as transition metal dichalcogenides and electron-hole bilayer systems~\cite{Kogar2017,Jauregui2019,Wang2019,Gupta2020NC,Ma2021,Sun2022NP,Jia2022NC,Gao2023,Song2023,Gao2024NC,Gao2024NP,Lian2024}, where reduced screening and enhanced Coulomb interactions facilitate exciton binding. In contrast, only a few bulk crystals have been proposed as excitonic insulators, with TiSe$_2$ and Ta$_2$NiSe$_5$ among the most studied~\cite{Pillo2000PRB,Kidd2002PRL,Rossnagel2002PRB,Cercellier2007PRL,Li2007PRL,Monney2009PRB,Monney2010PRB,Monney2011PRL,Watanabe2015PRB,Hedayat2019,Wakisaka2009,Seki2014PRB,Lu2017NC,Mor2017PRL,Sugimoto2018PRL,Bretscher2021,Kim2021,Kaneko2013PRB,Mazza2020PRL,Subedi2020PRM,Chen2023,Baldini2023PNAS,Watson2020PRR}. In TiSe$_2$, excitonic instability is closely intertwined with charge density wave (CDW) order, making it difficult to disentangle the intrinsic excitonic effects~\cite{Pillo2000PRB,Kidd2002PRL,Rossnagel2002PRB,Cercellier2007PRL,Li2007PRL,Monney2009PRB,Monney2010PRB,Monney2011PRL,Watanabe2015PRB,Hedayat2019}. Ta$_2$NiSe$_5$ undergoes a semiconductor-to-insulator transition attributed to exciton condensation~\cite{Wakisaka2009,Seki2014PRB,Lu2017NC,Mor2017PRL,Sugimoto2018PRL,Bretscher2021,Kim2021}. However, a concurrent structural phase transition has led to significant debate regarding its true nature~\cite{Kaneko2013PRB,Mazza2020PRL,Subedi2020PRM,Chen2023,Baldini2023PNAS,Watson2020PRR}. As a result, the experimental confirmation of exciton condensation in three-dimensional crystals remains challenging and controversial, highlighting the need for materials where exciton condensation can be disentangled from competing instabilities.

Recently, layered van der Waals compound Ta$_2$Pd$_3$Te$_5$ (TPT) has emerged as a promising platform for studying excitonic physics~\cite{Guo2021,Guo2022,Yao2024,Wang2021PRB,Wang2023NC,Li2024NC,Sun2023,Huang2024PRX,Zhang2024PRX,Hossain2023}. Theoretical calculations suggest that its monolayer structure could host an excitonic insulating state due to strong electron-hole interactions and a nearly zero-gap semimetallic band structure~\cite{Yao2024}. Previous studies have reported the observations of topological edge states~\cite{Wang2021PRB,Wang2023NC}. Subsequent angle-resolved photoemission spectroscopy (ARPES) measurements on bulk crystals have provided experimental evidence of a temperature-dependent band gap opening, indicating a transition from a Dirac semimetal at high temperatures to an insulating state at low temperatures~\cite{Huang2024PRX,Zhang2024PRX,Hossain2023}. Furthermore, unlike TiSe$_2$ and Ta$_2$NiSe$_5$, TPT shows no evidence of CDW or significant structural instability~\cite{Yao2024,Sun2023,Huang2024PRX,Zhang2024PRX,Hossain2023}, making it a cleaner system for probing exciton condensation. More importantly, current studies mainly focus on single-particle band renormalization, such as ARPES measurements, which only provide signatures from a single-particle perspective. Direct and definitive experimental evidence of many-particle collective excitations induced by exciton condensation has yet to be observed in optical spectroscopy.

\begin{figure*}[tb]
\includegraphics[width=2\columnwidth]{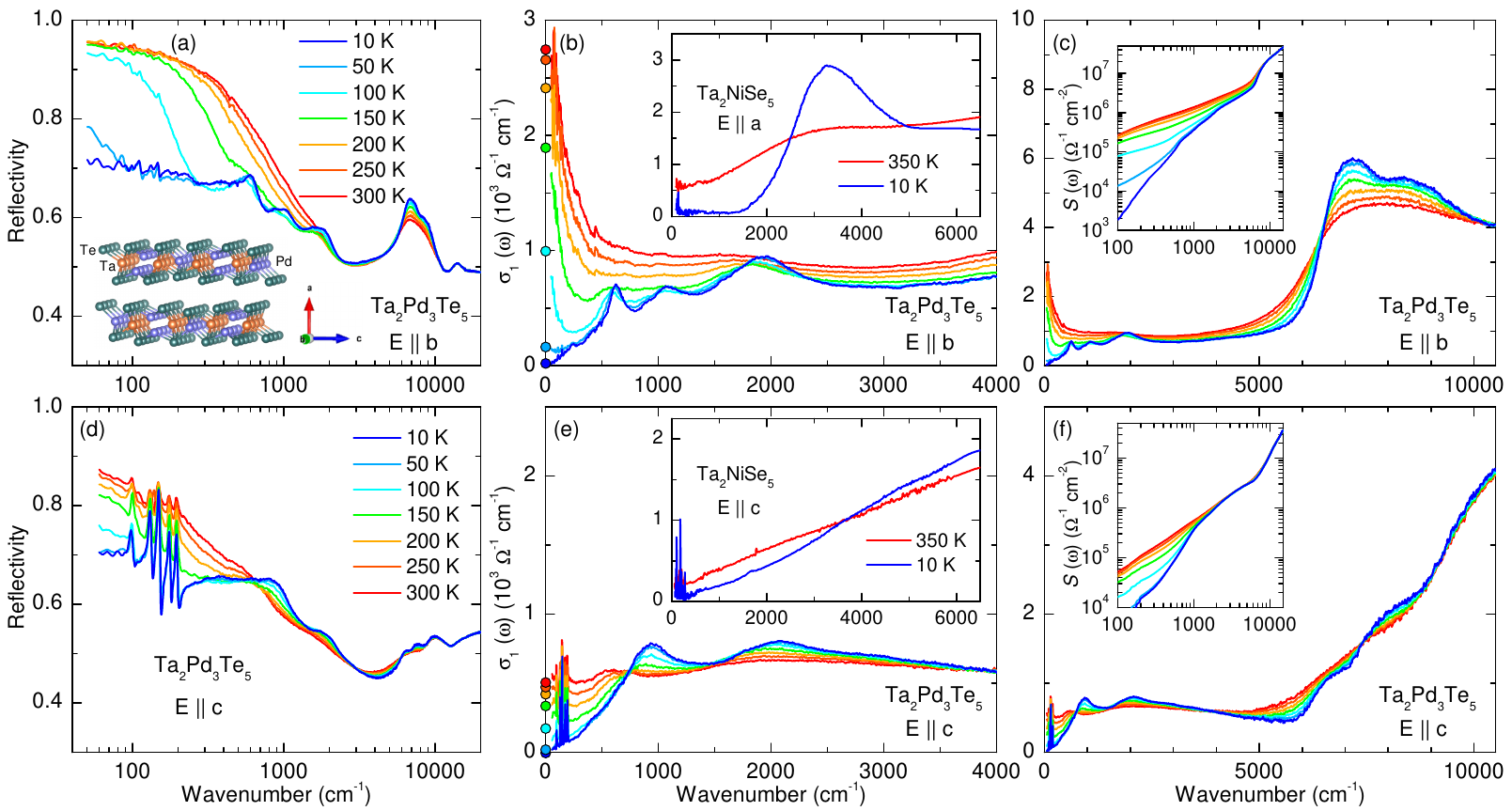}
\caption{(a) Temperature-dependent reflectivity of Ta$_2$Pd$_3$Te$_{5}$ along the $b$-axis. (b) and (c) Temperature-dependent optical conductivity along the $b$-axis up to 4\,000\icm\ and 10\,500\icm, respectively. Symbols on the $y$ axis of (b) denote dc conductivity values at corresponding temperatures from transport data. The inset of (c) shows the corresponding frequency-dependent spectral weight at different temperatures. (d--f) Corresponding plots for the data along the $c$-axis. The inset of (a) shows the crystal structure. The insets of (b) and (e) show the optical conductivity of Ta$_2$NiSe$_{5}$ along the $a$-axis and $c$-axis, respectively.}
\label{Fig1}
\end{figure*}

In this Letter, we present a comprehensive optical study of Ta$_2$Pd$_3$Te$_5$. Our results reveal a temperature-driven metal-insulator transition, accompanied by exceptionally narrow absorption features that significantly differ from those observed in previous excitonic insulator candidates. Based on the band structure calculations incorporating electron-hole interactions, we identify these sharp excitations as a hallmark signature of exciton condensation, originating from flat-band dispersions induced by excitonic order in Ta$_2$Pd$_3$Te$_5$.

%

Sample synthesis, experimental methods, and details of Drude-Lorentz analysis and theoretical calculations are provided
in the Supplemental Material~\footnote{See Supplemental Material at \url{http://link.aps.org/supplemental/xxx} for the details of sample synthesis, experimental methods, and details of Drude-Lorentz analysis and theoretical calculations.}.

%
%

Figures~\ref{Fig1}(a) and \ref{Fig1}(d) show the temperature-dependent reflectivity spectra of TPT along the $b$-axis and $c$-axis, respectively. The crystal structure and sample orientation are illustrated in the inset of Fig.~\ref{Fig1}(a). At room temperature, the $b$-axis reflectivity exhibits a typical metallic response with a plasma edge near 2\,000\icm\ and a prominent bump around 8\,000\icm\ due to strong interband transitions. Upon cooling, the plasma edge shifts to lower frequencies and vanishes below 50 K, signaling a metal-insulator transition. Along with the shift of plasma edge, three distinct peaks emerge at around 600, 1\,100, and 1\,800\icm. The $c$-axis reflectivity, $R_c(\omega)$, shows a similar temperature dependence but with lower values and stronger phonon features below 200\icm, indicating a weaker metallic response and reduced carrier screening. At low temperatures, $R_c(\omega)$ also reveals two more pronounced peaks at approximately 1\,000 and 2\,000\icm.

Figures~\ref{Fig1}(b) and \ref{Fig1}(e) display the temperature-dependent optical conductivity $\sigma_1(\omega)$ of TPT along the $b$-axis and $c$-axis, respectively, up to 4\,000\icm. At room temperature, the $b$-axis $\sigma_1(\omega)$ features a sharp Drude peak at zero frequency due to intraband excitations of free carriers, followed by a tail and a broad absorption around 1\,800\icm. As temperature decreases, the Drude peak gradually loses spectral weight (SW) and vanishes below 50 K, with $\sigma_1(\omega \rightarrow 0)$ decreasing in accordance with the inverse of the resistivity observed in transport (indicated by symbols on the $y$-axis). The absorption peak at 1\,800\icm\ shifts to around 2\,000\icm\ and narrows, while two additional narrow peaks emerge at around 600 and 1\,100\icm. A similar temperature dependence is observed along the $c$-axis, where the Drude SW diminishes and absorption peaks become prominent at around 1\,000 and 2\,000\icm. However, the Drude peak along the $c$-axis has much smaller SW, and only two peaks are visible at low temperatures. Notably, despite the Drude SW being quenched at 10 K along both axes, $\sigma_1(\omega)$ does not show a fully open gap but approaches zero or a small gap value. As discussed later, this behavior is attributed to the small indirect band gap in TPT, which prevents $\sigma_1(\omega)$ from vanishing entirely.

Figures~\ref{Fig1}(c) and \ref{Fig1}(f) show the temperature-dependent $\sigma_1(\omega)$ up to 10\,500\icm. Below 5\,000\icm, $\sigma_1(\omega)$ remains relatively flat, especially at high temperatures. As the frequency increases above 6\,000\icm, $\sigma_1(\omega)$ along both axes rises significantly, exhibiting a distinct double-peak structure at around 7\,000 and 9\,000\icm\ for the $b$-axis, and at around 6\,500 and 8\,000\icm\ for the $c$-axis. These peaks are indicative of strong interband absorptions. Notably, the absorption features along the $c$-axis are less pronounced compared to those along the $b$-axis, indicating the anisotropic nature of both the crystal and electronic structures in TPT.

\begin{figure}[tb]
\includegraphics[width=\columnwidth]{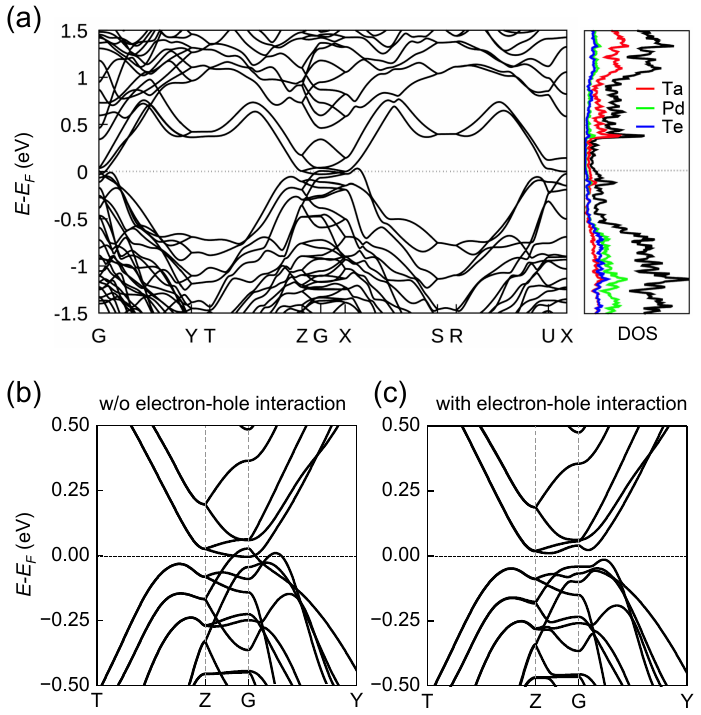}
\caption{(a) The mBJ band structure of bulk Ta$_2$Pd$_3$Te$_5$ and the corresponding density of states. (b) and (c) Band structures around the Fermi level without and with considering the electron-hole interaction, respectively.}
\label{Fig2}
\end{figure}

The metal-insulator transition or gap formation results in SW redistribution in $\sigma_1(\omega)$. We examine this redistribution with respect to both frequency and temperature using the quantity $S(\omega,T) = \int_{0}^{\omega} \sigma_{1}(\omega,T)d\omega$. The frequency-dependent $S(\omega,T)$ along the $b$-axis and $c$-axis at different temperatures is shown in the inset of Fig.~\ref{Fig1}(c) and Fig.~\ref{Fig1}(f), respectively. As the temperature decreases, the free carrier response is progressively suppressed and eventually quenched at low temperatures. For the $b$-axis, the lost SW does not fully recover within the energy scale of the weak low-energy absorptions below 5\,000\icm. Instead, it redistributes over a broad energy scale up to 10\,000\icm, corresponding to the strong high-energy absorptions. This suggests that a portion of the Drude SW is transferred to the high-energy interband transitions above 5\,000\icm. In contrast, for the $c$-axis, the frequency-dependent $S(\omega,T)$ merges at around 3\,000\icm, indicating that the lost SW from the Drude response is nearly recovered within the energy scale of the weak low-energy absorptions below 3\,000\icm.

\begin{figure*}[tb]
\includegraphics[width=2\columnwidth]{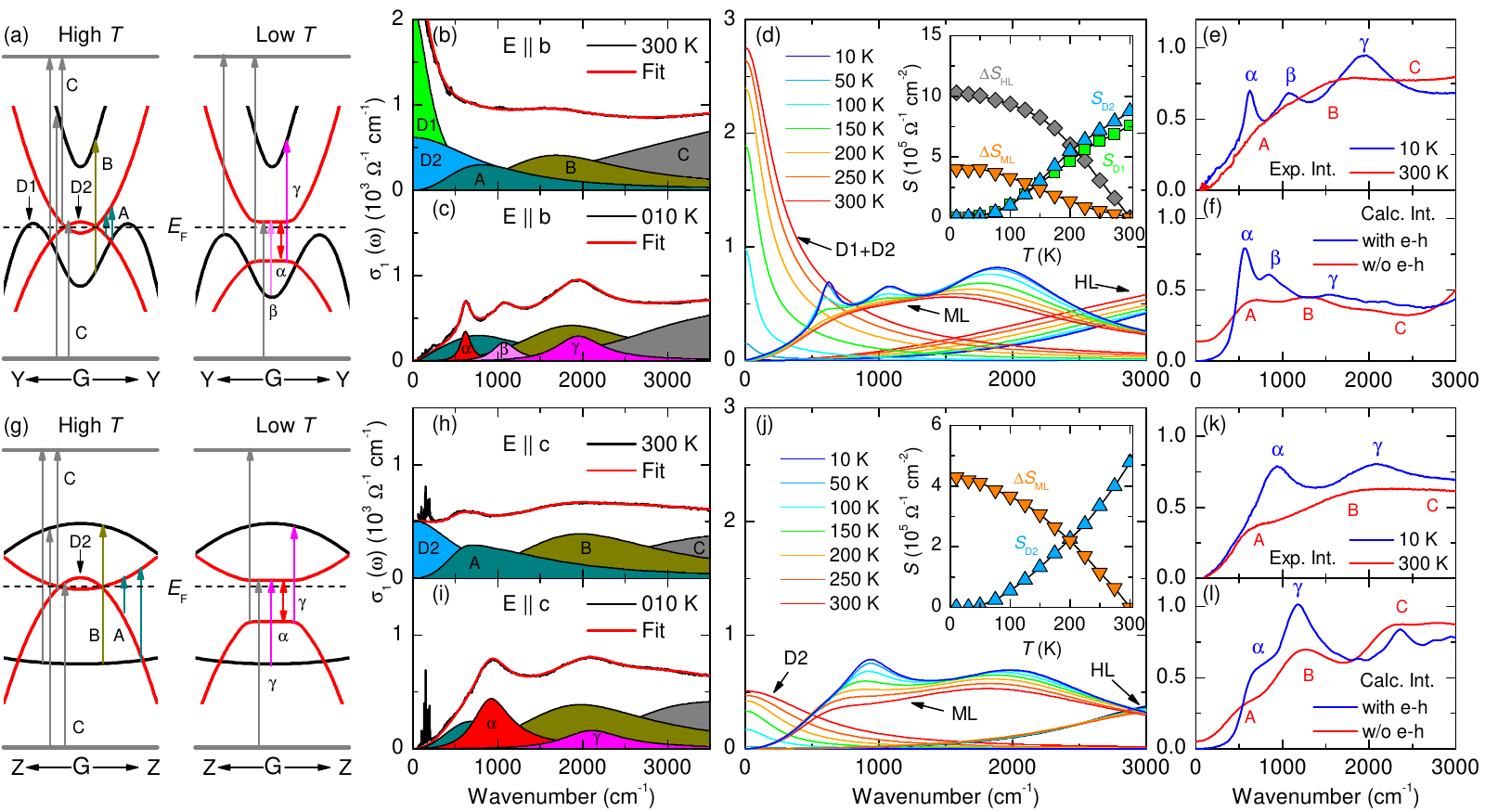}
\caption{(a) Schematic of the band structure of Ta$_2$Pd$_3$Te$_5$, without and with considering electron-hole interactions near the Fermi level along the Y--G--Y direction. (b) and (c) Drude-Lorenz fit of $\sigma_1(\omega)$ along the $b$-axis at 300 and 10 K, respectively. (d) Decomposition of $\sigma_1(\omega)$ at different temperatures along the $b$-axis. The inset shows the temperature-dependent spectral weight of different components: low-energy Drude response ($S_\mathrm{D}$), low-energy interband absorptions ($\Delta S_\mathrm{ML}$), and high-energy interband absorptions ($\Delta S_\mathrm{HL}$), where $\Delta S = S(T)-S(\mathrm{300~K})$. (e--f) A comparison of the interband $\sigma_1(\omega)$ obtained from (e) experimental measurements and (f) theoretical calculations. (g--l) Corresponding plots for the band structure along the Z--G--Z direction and the data along the $c$-axis.}
\label{Fig3}
\end{figure*}

To explore the nature of metal-insulator transition and low-energy excitations in TPT, we look into the band structure from DFT calculations. Previous studies have suggested a semimetallic band structure based on various calculation methods~\cite{Guo2021,Guo2022,Yao2024,Huang2024PRX}. The material retains its semimetallic nature even when the modified Becke-Johnson (mBJ) functional is applied. Figure~\ref{Fig2}(a) presents the mBJ band structure for bulk TPT, with calculation details provided in Supplemental Material~\footnotemark[1]. At the Fermi level ($E_\mathrm{F}$), the valence and conduction bands overlap, indicating a semimetallic state. From $-$0.5 to 0.5 eV, the bands show relatively large dispersion, while outside this range, the bands become denser with reduced dispersion. The corresponding density of states (DOS) is low in the $-$0.5 to 0.5 eV range but much higher outside this range. This band structure and DOS distribution align with the gross features of optical response, where $\sigma_1(\omega)$ shows weak absorptions below 6\,000 \icm\ ($\sim$ 0.8 eV), while stronger interband absorptions occur at higher energies. However, the semimetallic band structure cannot explain the observed insulating behavior and the emergence of sharp low-energy absorption features. To account for these, additional many-body interactions must be considered. Specifically, when electron-hole Coulomb interactions are included, recent calculations show that the bands near $E_\mathrm{F}$ in the monolayer TPT undergo significant renormalization or hybridization, leading to the opening of a gap~\cite{Yao2024,Huang2024PRX}. As shown in Figs.~\ref{Fig2}(b) and ~\ref{Fig2}(c), a comparison of the band structures of bulk TPT without and with considering the electron-hole interaction (calculation details are provided in Supplemental Material~\footnotemark[1]) reveals that the overlapped semimetallic bands opens a gap of approximately 0.08 eV around the G point. This gap provides a natural explanation for the insulating state observed at low temperatures. In fact, this gap was also confirmed in recent ARPES measurements, providing evidence of an excitonic insulating state in TPT. Note that, although a full gap opens near the G point, other bands remain close to $E_\mathrm{F}$, such as the top of the valence bands along the G--Y and X--S directions, resulting in a small indirect gap that prevents the full optical gap from being observed in $\sigma_1(\omega)$.

Based on the above discussions, we illustrate the band structures of TPT without and with electron-hole interactions near $E_\mathrm{F}$ along the Y--G--Y and Z--G--Z directions, shown in Figs.~\ref{Fig3}(a) and \ref{Fig3}(g), respectively. In the absence of interactions, both directions exhibit a pair of valence and conduction bands overlapped around $E_\mathrm{F}$ (red curves), with other nearby bands shown in black. Along the Y--G--Y direction (corresponding to the $b$-axis), an additional M-shaped band slightly crosses $E_\mathrm{F}$. Along the Z--G--Z direction (corresponding to the $c$-axis), another pair of valence and conduction bands lies slightly above and below $E_\mathrm{F}$. Bands farther from $E_\mathrm{F}$ are depicted by gray lines. At high temperatures, thermally excited carriers suppress the exciton condensation, allowing the non-interacting band structure to account for the observed optical response at high temperature. As shown in Fig.~\ref{Fig3}(b), we decompose the low-energy $\sigma_1(\omega)$ along the $b$-axis at 300 K using the Drude-Lorentz model (details are provided in Supplemental Material~\footnotemark[1]), which consists of two Drude terms for free carrier responses (labeled as D1 and D2) and several Lorentz terms for interband transitions (labeled as A, B, and C). Absorptions A and B are associated with transitions between the two pairs of bands near $E_\mathrm{F}$, while absorption C involves transitions to bands further away. A similar decomposition can be applied to the $c$-axis $\sigma_1(\omega)$. However, the $c$-axis shows only one Drude term from intraband excitations of the semimetallic bands. In contrast, the $b$-axis exhibits an additional Drude response due to the M-shaped band crossing $E_\mathrm{F}$. At low temperatures, electron-hole interactions dominate, leading to the spontaneous exciton formation and their subsequent condensation. This process flattens the dispersion of the top of the valence and the bottom of the conduction bands, opening a collective gap. This many-body insulating gap provides a framework for understanding the optical response of TPT at low temperatures. At 10 K, as shown in Figs.~\ref{Fig3}(c) and \ref{Fig3}(i), the decomposition of $\sigma_1(\omega)$ reveals that the Drude terms vanish in both directions. Several narrow peaks (labeled as $\alpha$, $\beta$, and $\gamma$) appear superimposed on the A and B absorptions. These peaks are associated with excitations of flat bands induced by exciton condensation, as indicated by the red and magenta arrows in Figs.~\ref{Fig3}(a) and \ref{Fig3}(f). Note that the $b$-axis exhibits an additional $\beta$ peak compared to the $c$-axis, owing to a better energy separation between the $\beta$ and $\gamma$ excitations.

Figures~\ref{Fig3}(d) and \ref{Fig3}(j) present the decomposition of $\sigma_1(\omega)$ at different temperatures along the $b$-axis and $c$-axis, respectively. To highlight the changes in SW, the spectrum is divided into three regions: Drude response, low-energy interband absorptions, and high-energy interband absorptions. As shown in the inset of Fig.~\ref{Fig3}(d), the redistribution of SW for the $b$-axis consists of two main parts: a portion of Drude SW that is transferred to the low-energy interband absorptions ($\Delta S_\mathrm{ML}$), and additional Drude SW that is transferred to high-energy interband absorptions ($\Delta S_\mathrm{HL}$). This is attributed to the fact that the M-shaped band along the $b$-axis crosses $E_\mathrm{F}$ at high temperatures, but sinks completely below $E_\mathrm{F}$ at low temperatures, causing the Drude SW to be lost and transferred to high-energy interband transitions. In contrast, for the $c$-axis, the redistribution of SW is mainly between the Drude response and the low-energy interband absorptions, as evident in the inset of Fig.~\ref{Fig3}(j).

To further underscore the impact of exciton condensation on low-energy excitations, we present a direct comparison of the low-energy interband $\sigma_1(\omega)$ with and without the electron-hole interaction. As shown in Figs.~\ref{Fig3}(e--f) and Figs.~\ref{Fig3}(k--l), both experimental data and theoretical calculations highlight the emergence of sharp features ($\alpha$, $\beta$, and $\gamma$) associated with flat bands. In particular, the $\alpha$ peak corresponds to quasiparticle excitations across the collective gap induced by exciton condensation. Along the $b$-axis, the $\alpha$ peak appears at 620\icm\ (77 meV) with a narrow half-width of only 150 cm$^{-1}$ (19 meV) at 10 K, well reflects the excitation characteristics of flat bands. This behavior contrasts remarkably with other excitonic insulator candidates, such as TiSe$_2$ and Ta$_2$NiSe$_5$. For instance, in Ta$_2$NiSe$_5$~\cite{Lu2017NC}, the absorption peak is located at $\sim$ 3\,000\icm\ (0.4 eV) with a broad half-width of nearly 2\,000\icm\ along the $a$-axis, while no absorption peak is seen along the $c$-axis, as shown in the insets of Figs.~\ref{Fig1}(b) and \ref{Fig1}(e). Similarly, in TiSe$_2$~\cite{Li2007PRL}, excitonic instability induces a CDW gap, with an absorption peak also around 3\,000\icm\ and a half-width of $\sim$ 2\,000\icm. The $\alpha$ peak in TPT is at least one order of magnitude narrower, more clearly representing intrinsic excitations and coherence of exciton condensation, rather than excitonic instability coupled with competing orders that lead to the charge-ordered state like CDW in TiSe$_2$~\cite{Pillo2000PRB,Kidd2002PRL,Rossnagel2002PRB,Cercellier2007PRL,Li2007PRL,Monney2009PRB,Monney2010PRB,Monney2011PRL,Watanabe2015PRB,Hedayat2019} or the structural order-induced gap opening in Ta$_2$NiSe$_5$~\cite{Kaneko2013PRB,Mazza2020PRL,Subedi2020PRM,Chen2023,Baldini2023PNAS,Watson2020PRR}. As the temperature increases, the sharp peak features broaden and gradually vanish. Specifically, a substantial reduction in the intensity of the $\alpha$ peak along the $b$-axis occurs at 200 K, significantly lower than its respective peak energy (620 cm$^{-1}$ $\sim$ 890 K). This behavior cannot be explained by a single-particle gap but instead suggests a transition from an excitonic insulating state to a metallic state, driven by thermal excitation disrupting the coherence of exciton condensation. These findings strongly support the presence of exciton condensation at low temperatures in bulk TPT. Additionally, the differences in half-width and peak position of the $\alpha$ peak along the $b$-axis and $c$-axis further indicate anisotropic excitonic dynamics in TPT.

%
%
In summary, we observed a distinct metal-insulator transition in the optical conductivity of bulk Ta$_2$Pd$_3$Te$_5$, accompanied by several sharp absorption peaks at low temperatures. These phenomena are consistent with theoretical predictions of a many-body gap driven by electron-hole interactions. The absence of competing orders, along with the narrow intrinsic excitonic features, distinguishes Ta$_2$Pd$_3$Te$_5$ from previous candidates, providing strong evidence for exciton condensation. Therefore, Ta$_2$Pd$_3$Te$_5$ offers a promising platform for exploring the novel states of matter and holds potential for applications in exciton-based devices.

%
%
\begin{acknowledgments}
This work was supported by the National Key Research and Development Program of China (Grants No. 2022YFA1403900, No. 2024YFA1408301, No. 2023YFA1406002, No. 2024YFA1408400, No. 2022YFA1403800, and No. 2021YFA1400401), the National Natural Science Foundation of China (Grant No. 12274442, No. 12374155, No. 12188101, and No. U22A6005), and the Center for Materials Genome.
\end{acknowledgments}

%
%

\end{document}